# Electronic diffusion in a normal state of high-Tc cuprate $YBa_2Cu_3O_{6+x}$

Jure Kokalj[a,b]

[a]Jožef Stefan Institute, 1000 Ljubljana, Slovenia; and [b]University of Ljubljana, Faculty of Civil and Geodetic Engineering, 1000 Ljubljana, Slovenia.

* Jure Kokalj.

**Email:** jure.kokalj@ijs.si

**Abstract**

The bad metallic phase with resistivity above the Mott-Ioffe-Regel limit, which appears also in cuprate superconductors, was recently understood by cold atom and computer simulations of the Hubbard model via charge susceptibility and charge diffusion constant. However, since reliable simulations can be typically done only at temperatures above the experimental temperatures, the question for cuprate superconductors is still open. This paper addresses this question by resorting to heat transport, which allows for the estimate of electronic diffusion and it further combines it with the resistivity to estimate the charge susceptibility. The doping and temperature dependencies of diffusion constant and charge susceptibilities are shown and discussed for two samples of $YBa_2Cu_3O_{6+x}$. Results indicate strongly incoherent transport, mean free path corresponding to the Mott-Ioffe-Regel limit for the underdoped sample at temperatures above ~200 K and significant effect of the charge susceptibility on the resistivity.

**Main Text**

Understanding the resistivity in the normal state of high-Tc cuprates poses one of the central challenges in strongly correlated systems for almost four decades and could be a crucial step for understanding the superconductivity. Resistivity shows unusually high values reaching well above the proposed Mott-Ioffe-Regel (MIR) limit at high temperatures (e.g. above 300 K for low hole dopings in $YBa_2Cu_3O_{6+x}$). Such bad-metallic behavior was recently understood with cold atoms (1) and computer (2,3) simulations of Hubbard model in terms of strongly suppressed charge susceptibility $\chi_c$, which results in large resistivity $\rho$ via the Nernst-Einstein relation $\rho = 1/(D_c \chi_c)$. It was further shown that besides the charge diffusion constant $D_c$, which decreases with $T$ and saturates at very high $T$, also $\chi_c$ decreases with $T$ and importantly affects the $T$ dependence of $\rho$. Model simulations correspond to very high $T$ and merely touch cuprate's experimental high-$T$ regime. This raises an important question. Is the behavior observed in simulations for high $T$, e.g., the importance of the $T$-dependence of $\chi_c$ stretched into the experimental lower $T$ regime for cuprates? This seems plausible, since for some dopings $\rho$ shows no qualitative change from highest $T$ all the way down to the superconducting transition temperature $T_c$. This question could be answered with data either on $\chi_c$ or $D_c$, but both are still elusive for experimental probing. Here we resort to another line of research with many remarkable discoveries for cuprates: the study of thermal properties. Measured specific heat $c$ and its electronic part $c_{el}$ stay robust with practically no corrections since the early measurements (4). Heat conductivity $\kappa$ offered an early support for the d-wave pairing in cuprates (5) and revealed increased heat conduction below $T_c$ triggering hot debate (6,7) on its origin: increased mean-free-path of either phonons or non-superconducting electrons. The latter was strongly supported by the increased infrared conductivity (8,9). For the insulating parent compound, the magnonic $\kappa$ was determined as the difference between the measured in-plane and out-of-plane conductivities allowing the extraction (10) of very long magnonic mean free paths (longer than 100 unit cells). Introductions of impurities (9,10), defects (11) and magnetic fields (10) were used to reveal the conduction and scattering mechanisms. Furthermore, directly measured heat diffusion constant was discussed in terms of Planckian dissipation and a soup like mixture of electrons and phonons (12).

**Results and Discussion**

Heat transport is special in a sense that all quantities in its Nernst-Einstein equation $\kappa = D_Q c$ can be independently measured - even the heat diffusion constant $D_Q$ (12). The challenge however is the separation of phononic $\kappa_{ph}$ and electronic $\kappa_{el}$ contributions to the total conductivity $\kappa = \kappa_{ph} + \kappa_{el}$. Takenaka et al. (9) carefully analyzed the O and Zn doping dependence of $\kappa$ for YBa$_2$Cu$_3$O$_{6+x}$ (YBCO) and obtained good estimates for $\kappa_{el}$ (Fig. 1A). We use these data, together with the data for $c_{el}$ (Fig. 1B) from Loram et al. (4) to estimate the electronic heat diffusion constant $D_{Q,el}$ via the Nernst-Einstein relation $D_{Q,el} = \kappa_{el}/c_{el}$. The resulting $D_{Q,el}$ for two measured compounds with $T_c \sim$ 60 K (x ~ 0.68) and $T_c \sim$ 90 K (x ~ 0.93) in the non-superconducting regime is shown in Fig. 1C. $D_{Q,el}$ decreases with increasing *T* as expected, since it is related to the mean free path $l$ via $D_{Q,el} = vl/2$ and $l$ normally decreases with increasing *T* due to increasing scattering. Here $v$ is a mean quasiparticle velocity. By using its estimate $v = 2.15 \times 10^5$ m/s (12) and setting $l$ to minimal value ($l \sim a$ with $a$ being a lattice constant) one obtains a MIR limit for $D_{Q,el}$ (Fig. 1C). Remarkably, $D_{Q,el}$ for the $T_c \sim$ 60 K compound decreases to such value already at relatively low $T \sim$ 200 K and shows signs of saturation close to MIR value at higher $T$. This reveals strongly incoherent behavior in such regime. From value of $D_{Q,el}$ and $v$ one can estimate $l$ (right y axis in Fig. 1C). This average value of $l$ is small and reaches $l \sim 6a$ at lowest $T \sim T_c$. $T$ dependence of $D_{Q,el}$ can be related to the $T$ dependence of scattering time $\tau$ via $D_{Q,el} = v^2\tau/2$. In Fig. 1D we show that for $T_c \sim$ 60 K compound the behavior at lowest $T$ is most consistent with Fermi liquid (FL) like scattering $1/\tau \propto T^2$, advocated, e.g., in Ref. 13, while the $T_c \sim$ 90 K compound shows more $1/\tau \propto T$ like behavior. We show in Fig. 1D also the comparison to the suggested (14) lower bound on diffusion $D_{p.b.} = \alpha\hbar v^2/(2k_B T)$ obtained from a Planckian timescale $\tau = \alpha\hbar/(k_B T)$. $D_{p.b.}$ is shown for a numerical prefactor $\alpha$ of the order unity set to $\alpha = 1$. $D_{Q,el}$ is much smaller and the consistency with $D_{p.b.}$ for a $T_c \sim$ 90 K compound requires $\alpha \sim 0.2$, while smaller $\alpha \sim 0.05$ is needed for consistency with $T_c \sim$ 60 K compound.

By having values for $D_{Q,el}$, we can make a step towards estimating $\chi_c$. $D_c$ and $D_{Q,el}$ typically do not behave the same, but they are related. It was found that within a dynamical mean field theory (15) they are at low $T$ related by $D_c = fD_{Q,el}$, with a factor $f = 1/z^2$ and $z$ being a quasiparticle weight. By assuming a constant factor $f$ in regime of measured data and setting it to an approximate value $f = \frac{1}{0.4^2}$, we obtain a sensible approximation for $D_c$. This can be used together with the data on resistivity (9) (Fig. 2A) to calculate $\chi_c$ via the Nernst-Einstein relation $\chi_c = 1/(\rho D_c)$. The resulting $\chi_c$ is shown in Fig. 2B. It is compared to the theoretical estimate $\chi_c \approx e_0^2 z g_0$, with $g_0$ being a non-interacting density of states. $\chi_c$ is larger than theoretical estimate and in addition shows some *T* dependence. A clear cusp is seen for $T_c \sim$ 60 K compound at $T \sim$ 120 K presumably corresponding to the charge density-wave (CDW) phase transition (16). More relevant for our discussion is its *T* and doping dependence above such *T*. While both compounds show decreasing $\chi_c$ with increasing *T*, the decrease for $T_c \sim$ 60 K compound is stronger. From 200 *K* to 300 K its $\chi_c$ decreases by about 15%, which is comparable to the decrease of $D_c$ for about 20% (together they give an increase of $\rho$ for about 50%). This is similar to the behavior observed in the Hubbard model (1). *T* dependence of $\chi_c$ therefore plays an important role for $\rho$ in YBCO and may turn out to be even greater at higher *T* as $D_c$ is expected to be saturating while $\rho$ increases further. This agrees well with the picture from the optical conductivity (17), in which the increase of resistivity with increasing *T* in the bad metallic phase is due to the transfer of low frequency spectral weight to higher frequencies, while the mean free path is saturating. It is interesting to note, that with decreasing doping, $\chi_c(T \sim 150\ \text{K})$ is increasing, which is consistent with the phase separation tendency, for which $\chi_c$ diverges at some point in the extended phase diagram (18). We also note that the Nernst-Einstein relation is applicable also to the pseudogap phase as it relies (2,14) only on the normal behavior with some scattering or current relaxation mechanism leading to the finite diffusion constant.

One should keep in mind, that the analysis here is based on experimental data, which have some degree of uncertainty and that the properties are averaged over all carriers with no differentiation, e.g., nodal vs. anti-nodal Fermi surface parts. However, the main outcomes like strongly incoherent electronic transport close to expected MIR saturation and the first estimate of $\chi_c$ suggesting notable $T$ and doping dependence could withstand future tests. The presented behavior and insight also hint on possible better understanding and calls for further experimental efforts to pin down at least one additional quantity, either charge diffusion constant $D_c$ or charge susceptibility $\chi_c$.

**Materials and Methods**

Some details on calculations can be found in the SI Appendix.

All the data are taken from references and results calculated as explained in the text.

**Acknowledgments**

Author thanks J. R. Cooper for sharing specific heat data and J. Mravlje for helpful discussions and acknowledges the support from the Slovenian Research and Innovation Agency (ARIS) under Grant No. P1-0044.

12. J. Zhang *et al.*, Anomalous thermal diffusivity in underdoped YBa$_2$Cu$_3$O$_{6+x}$. Proc. Natl. Acad. Sci. **114**, 5378 (2017).
13. N. Barišić *et al.*, New J. Phys. **21**, 113007 (2019).
14. S. A. Hartnoll, Nat. Phys. **11**, 54 (2015).
15. M. Ulaga *et al.*, Thermal conductivity and heat diffusion in the two-dimensional Hubbard model. Phys. Rev. B **106**, 245123 (2022).
16. S. Blanco-Canosa *et al.*, Resonant x-ray scattering study of charge-density wave correlations in YBa$_2$Cu$_3$O$_{6+x}$. Phys. Rev. B **90**, 054513 (2014).
17. N. E. Hussey *et al.*, Universality of the Mott-Ioffe-Regel limit in metals. Philos. Mag. **84**, 2847 (2004).
18. T. Misawa and M. Imada, Phys. Rev. B **90**, 115137 (2014).

**Figures**

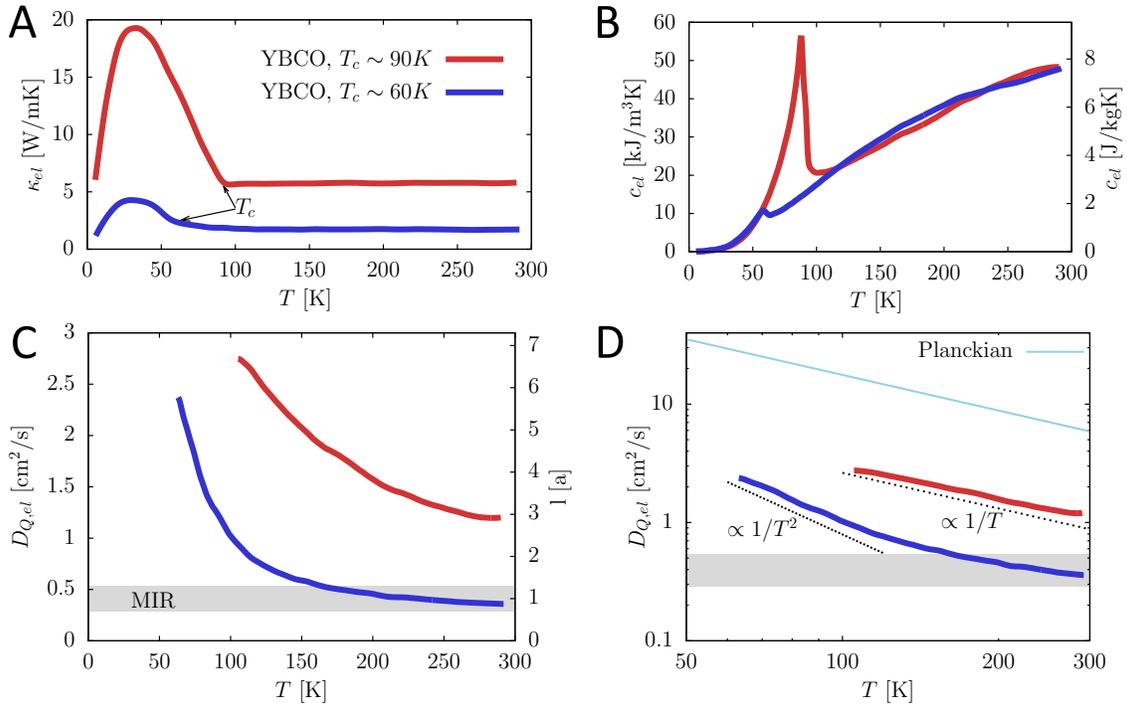

**Figure 1.** A) Electronic heat conductivity $\kappa_{el}$ for two YBCO samples ($T_c \sim 60$ K and $T_c \sim 90$ K) from Ref. 9. Transition temperatures $T_c$ are indicated. B) Electronic specific heat as measured for similar samples from Ref. 4. C) Calculated electronic heat diffusion constant $D_{Q,el}$ for two samples showing decrease with *T* and for $T_c \sim 60$ K sample also values and tendency for saturation close to the MIR limit at higher *T*. Due to some uncertainty in definition of MIR limit, it is shown with 30% range from $l = a$. Right *y*-axis shows estimated mean free path. D) $D_{Q,el}$ in logarithmic scale compared with $1/T^2$, $1/T$ and suggested Planckian lower bound for a prefactor $\alpha = 1$.

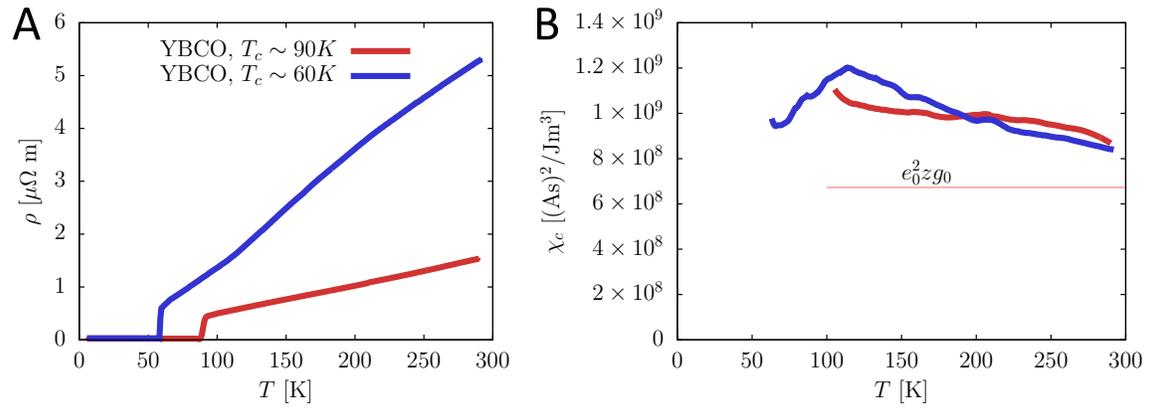

**Figure 2.** A) Resistivity $\rho$ for the same YBCO samples as in Fig. 1. B) Via the Nernst-Einstein relation deduced charge susceptibility $\chi_c$ and its comparison with theoretical estimate. $\chi_c$ shows notable *T* and doping dependence, as well as cusp at presumably CDW transition.

Supplementary Information:

**Extended materials and methods**

$\chi_c = e_0^2 dn/d\mu$ with $n$ being a density of electrons and $\mu$ the chemical potential. $\chi_c$ can be for non-interacting particles related to the density of states at the Fermi energy, while it is renormalized in the presence of interactions (1,2). First approximation gives $\chi_c \approx e_0^2 z g_0$ and we estimate the non-interacting density of states as $g_0 = \sqrt{\frac{2}{\pi}} \frac{2}{\hbar c a v_0}$ corresponding to an isotropic 2D Fermi surface. We use bare band velocity $v_0 = 5.2 \times 10^5$ m/s estimated from LDA band dispersions (3), lattice constants $c = 11.68$ Å and $a = 3.82$ Å (4), half-filled band and two Cu-O planes within $c$. For unit cell volume the lattice constant $b = 3.89$ Å (4) is also used and mass density 6.3 g/cm$^3$ for transformation of units between left and right *y* axes in Fig. 1B.